# Energy-Dependent Electron-Electron Scattering and Spin Dynamics in a Two Dimensional Electron Gas.

W.J.H.Leyland[a,b], R.T.Harley[a], M.Henini[c], A.J.Shields[d], I.Farrer[b] and D.A.Ritchie[b]

[a]*School of Physics and Astronomy, University of Southampton SO17 1BJ, UK*
[b]*Cavendish Laboratory, University of Cambridge, Madingley Road, Cambridge CB3 0HE, UK*
[c]*School of Physics and Astronomy, University of Nottingham NG7 4RD, UK*
[d]*Toshiba Research Europe Ltd, Cambridge CB4 4WE, UK*

**Abstract**

Measurements of spin dynamics of electrons in a degenerate two dimensional electron gas, where the Dyakonov-Perel mechanism is dominant, have been used to investigate the electron scattering time ($\tau_p^*$) as a function of energy near the Fermi energy. Close to the Fermi energy the spin evolution is oscillatory, indicating a quasi-collision-free regime of spin dynamics. As the energy is increased a transition to exponential, collision-dominated, spin decay occurs. The frequency and the value of $\tau_p^*$ are extracted using a Monte Carlo simulation method. At the Fermi energy $\tau_p^*$ is very close to the ensemble momentum relaxation time ($\tau_p$) obtained from the electron mobility. For higher energies $\tau_p^*$ falls quadratically, consistent with theoretical expectations for the onset of electron-electron scattering which is inhibited by the Pauli principle at the Fermi energy.

**Introduction**

Investigation of the electron spin dynamics in semiconductors can reveal detail of both the spin-orbit interaction and of electron scattering mechanisms. Here we exploit this to study experimentally the energy dependence of electron-electron scattering near the Fermi level of a two-dimensional electron gas.

This is possible because in III-V semiconductors and especially in quantum wells the Dyakonov-Perel mechanism [1,2] is usually the most important for relaxation of a non-equilibrium electron spin population. In this case reorientation of the spins is caused by spin-orbit interaction which can be represented as an effective, momentum-dependent magnetic field or precession vector $\Omega_\mathbf{k}$ acting on an electron's spin. Scattering which randomises the momentum of an electron thus randomises the precession and if the scattering rate ($1/\tau_p^*$) exceeds the precession frequency, as is normal, the spin evolution is an exponential relaxation. In this way scattering which does not result directly in spin-flips produces spin relaxation. The spin relaxation rate along an axis $i$ is [1,3]

$$1/\tau_{s,i} = <\Omega_\perp^2> \tau_p^* \tag{1}$$

where $<\Omega_\perp^2>$ is the average square of the component of the precession vector in the plane perpendicular to $i$.

In the (less usual) case of weak momentum scattering an electron spin may precess through several complete cycles before undergoing momentum scattering [4,5]. The approach to equilibrium of the spin population may then be oscillatory provided its precession frequency is sufficiently homogeneous, for example if the spin polarized electrons are all close to the Fermi momentum in a two dimensional electron gas (2DEG) [5,6].

In recent papers [5,6,7] we have described optical investigations of electron spin dynamics near the Fermi energy ($E_F$) in a series of high mobility 2DEGs confined in GaAs/AlGaAs quantum wells at temperatures from 1.5K to 300K. This demonstrated the transition from the weak scattering regime at low temperatures to the strong scattering regime at high temperatures. At low temperatures we have observed oscillatory spin dynamics allowing separate determination of the momentum scattering rate and the spin-orbit interaction as a function of quantum well width [5,7]. The scattering rate ($1/\tau_p^*$) was found to be close to the ensemble momentum relaxation rate ($1/\tau_p$) given by measurement of the mobility. At high temperatures we observed exponential spin relaxation with a rate which fell rapidly to a minimum value in the region of the Fermi temperature ($E_F/k_B$) [6]. We showed that this behaviour, a clear example of the motional narrowing (or slowing) characteristic of the Dyakonov-Perel mechanism [1] (see eq. 1), is caused by the onset of strong electron-electron scattering as the 2DEG moves away from full degeneracy [8]. We found that, in general, the total scattering rate is given by

$$1/\tau_p^* = 1/\tau_p + 1/\tau_{ee} \tag{2}$$

where $1/\tau_{ee}$ is the electron-electron scattering rate.

In this paper we describe spin-dynamic measurements of the energy dependence of the scattering. By working at low temperatures we separate the precession and scattering components of the spin dynamics and we vary the electron energy with respect to the Fermi energy at fixed values of temperature by varying the photon energy which is used to inject the spin-polarised electrons into the 2DEG. The scattering rate $1/\tau_p^*$ increases rapidly with energy in reasonable agreement with the theoretical expectation that the contribution from $1/\tau_{ee}$ vanishes for electrons at the Fermi surface due to the Pauli principle and increases approximately quadratically with increase of energy [9,10,11]. Our results are comparable with those obtained from four-wave-mixing measurements by Kim et al. [12].

**2) Experimental details.**

The sample used for these measurements was a (001)-oriented one-side *n*-modulation doped single GaAs/Al$_{0.35}$Ga$_{0.65}$As quantum well with nominal width 20 nm. It is one of the series used in our previous studies [5,6,7], T539, and was chosen because it has the highest electron mobility at 5 K corresponding to ensemble momentum relaxation time $\tau_p$ = 27 ps [6]. The electron confinement energy $E_{1e}$ = 10.2 meV and Fermi energy $E_F$ = 6.2 meV were determined using photoluminescence (PL) and photoluminescence excitation (PLE) spectroscopy [6]. The spin-dynamics of the 2DEG was investigated using the picosecond-resolution pump-probe optical reflection technique we have described earlier [13]. The circularly-polarized pump beam intensity was typically 0.5 mW focused to a 60 micron diameter spot giving an estimated photoexcited spin-polarized electron density 5x10$^9$ cm$^{-2}$, very much less than the unpolarised electron concentration in the 2DEG; the probe power density was 25% of the pump. The pump and probe photon energies were degenerate and could be tuned to energies spanning the Fermi energy of the 2DEG. The technique measures both the time-evolution of the probe polarization rotation, $\Delta\theta$, which is proportional to the component of electron spin-polarisation along the growth axis, $<S_z>(t)$, and also the change of reflected intensity $\Delta R$, which gives the photo-excited population. Over the timescale of the decay of $\Delta\theta$ we found that $\Delta R$ was essentially constant so that the observed evolution of $\Delta\theta$, and therefore of $<S_z>(t)$, is not significantly affected by recombination of the spin-polarised electrons. The spectral resolution of these measurements is set by the pulse duration of pump and probe pulses and is ~ 0.8 meV. For measurements at 5 K the sample was mounted in a liquid helium flow cryostat surrounded by cold gas whereas for 1.5 K measurements it was immersed in superfluid helium. Thermal rounding of the Fermi surface is ~ 0.5 meV at 5 K and ~ 0.15 meV at 1.5 K.

**3) Results.**

Fig. 1 shows examples of $\Delta\theta$ (ie $<S_z>(t)$) signals at 1.5 K and at 5 K measured at several different values of excitation energy expressed in the form $E/E_F$ where $E$ is measured from the bottom of the *n*=1 conduction sub-band. The photon energy required to excite an electron to the Fermi level of the 2DEG was taken to be the half-intensity point of the onset of PLE, 1.5241 eV at 5 K (see Fig. 3b) and the zero of the energy scale is taken 6.2 meV lower, 1.5179 eV. At 1.5 K the spin evolution was a damped oscillation for all energies investigated, as indicated for example by the upper trace in Fig. 1, whereas at 5 K the evolution was found to change progressively from oscillatory to monotonic and eventually exponential as the energy was increased (see lower traces of Fig. 1).

It can be seen in Fig.1 that the sign of the signal inverts just above $E_F$. This is a feature of the nonlinear reflectivity technique which we have found in all the samples we have investigated and is discussed more fully in a previous paper [7]. Although the magnitude and sign of the signals are functions of energy, the extracted values of precession frequency and scattering rate are unaffected by the phase inversion. Maximum signal strength was observed adjacent to the phase inversion, on each side, falling to the noise level ~ 5 meV above the Fermi energy.

Fig. 2 shows results of a Monte Carlo simulation of the data for a precession frequency $\Omega_{ex}$= 0.063 radians/ps and a range of values of $\tau_p^*$. The calculations were performed for a population of $10^4$ electrons injected at time zero with spins oriented along the growth axis and momenta distributed at random on the Fermi surface. Evolution of the $z$-component of the total spin of this ensemble was calculated in time steps $\delta t = 1$ ps. During each step each electron spin was subject to a precession vector determined by its wavevector, assuming the Dresselhaus form of spin-splitting. At the end of each step it was subject to a possibility of elastic scattering $\delta t/\tau_p^*$ to a new randomly chosen wavevector giving a new precession vector. The simulation reproduces the experimental behaviour very well and enables extraction of reliable values of precession frequency and the momentum scattering time $\tau_p^*$. We have described our procedure for this in previous papers [5,7]. In the simulations the boundary between oscillatory and monotonic spin evolution occurs at $\Omega_{ex}\tau_p^* = \frac{1}{2}$.

The extracted precession frequency was found to be constant within experimental uncertainties at 0.063±0.006 radians/ps over the energy and temperature range investigated whereas the scattering time which is closely correlated with the damping of the observed signal varied strongly. Fig.3a shows the values of $\tau_p^*$ as a function of electron energy; there is a plateau in the vicinity of the Fermi energy and a very rapid fall with increasing energy. The solid points indicate oscillatory spin evolution whereas monotonic decay is indicated by open symbols.

**4) Discussion and conclusions**

First consider the observed transition from oscillatory to monotonic spin dynamics as indicated by the solid and open symbols in Fig. 3a. The boundary separating shaded and unshaded regions in Fig. 3a is a plot of $1/(2\Omega_{ex})$ where $\Omega_{ex}$ is the calculated value of precession frequency at the excitation energy taking the value at the Fermi energy to be $\Omega_F = 0.063$ radians/ps and assuming $\Omega_{ex} \sim E^{1/2}$. The curve corresponds to the condition $\Omega_{ex}\tau_p^* = \frac{1}{2}$, indicated by the Monte Carlo simulations to be the boundary between oscillatory and monotonic spin evolution. As expected it nicely separates the regions where oscillatory and monotonic spin evolution are observed experimentally.

Now consider the observed energy and temperature dependence of the electron momentum scattering time $\tau_p^*$ (Fig.3a). Since the contribution from the ensemble momentum relaxation rate $1/\tau_p$, as indicated by the measured mobility, is essentially constant in this region, the variation must be due primarily to the electron-electron scattering rate $1/\tau_{ee}$. The latter has been considered theoretically by several authors. Analytic expressions for the temperature dependence of the scattering rate at energy $E_F$ and for the energy dependence at $T = 0$ were derived in refs. 9 and 10. Comparing these formulae with numerical calculations of the combined effect of energy and temperature variation, Jungwirth and MacDonald [11] have concluded that for $T/T_F \leq$ 0.1 and energies $\zeta \equiv (E-E_F)/E_F \leq 0.1$ a sum of the analytic expressions is a very good approximation. For our sample $T_F = 72$K so these condition hold and we expect

$$1/\tau_{ee}^*(\zeta,T) = 1/\tau_{ee}(\zeta,0) + 1/\tau_{ee}(0,T) \qquad (3)$$

where

$$1/\tau_{ee}(\zeta,0) = A.E_F\zeta^2\left[1.54 - \ln\zeta\right] \qquad (4)$$

and

$$1/\tau_{ee}(0,T) = A.\frac{\pi^2}{2}E_F\left(\frac{k_BT}{E_F}\right)^2\left[0.96 - \ln\left(\frac{k_BT}{E_F}\right)\right] \qquad (5)$$

*A* is a constant which specifies the electron-electron scattering strength. The dotted and solid curves in Fig. 3a are calculated values of $\tau_p^*$ at *T*=1.5K and at 5K respectively obtained using eqs. 3, 4 and 5 and with $\tau_p$ set at the value 27 ps indicated by our mobility measurements. The value of *A* was chosen to be 0.4 to give a fit to the data. This value of the scattering strength is consistent with that used in our analysis of the full temperature dependence of the spin relaxation time described in our previous paper [6]. The agreement between the calculation and experiment is reasonably good indicating the correctness of the approach.

In conclusion, we have used measurements of spin dynamics to investigate the energy and temperature dependence of electron-electron scattering in the vicinity of the Fermi energy in a quasi-degenerate two dimensional electron gas. The scattering time for electrons at the Fermi energy is close to the momentum scattering time given by a measurement of the electron mobility and decreases quadratically as the energy is increased. The form and magnitude of the decrease is consistent with additional scattering due to electron-electron interaction in accordance with theoretical expectations. The observed spin dynamics show a transition from oscillatory to exponential time evolution as the electron energy increases which is very well accounted for by a Monte Carlo simulation based on the Dyakonov-Perel mechanism of spin dynamics.

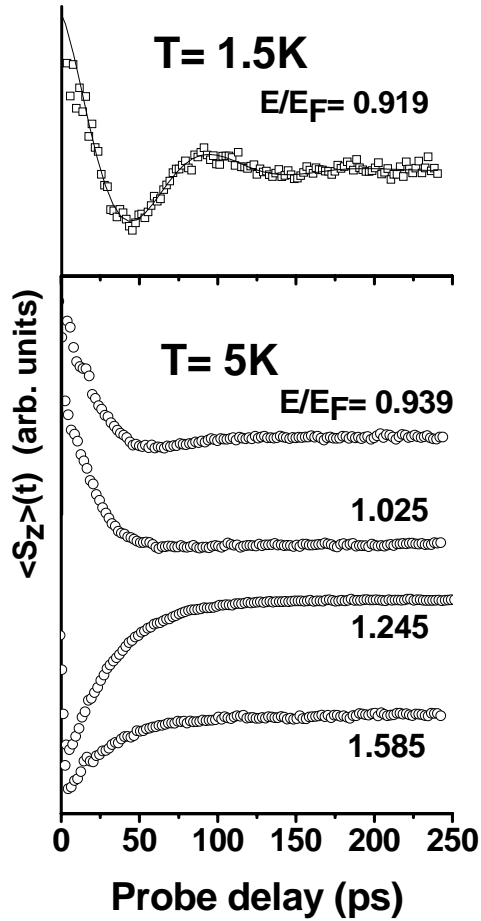

**Figure 1**  Examples of $\Delta\theta$ (ie $<S_z>(t)$) signals at 1.5K and at 5K for a range of electron energies relative to the Fermi energy. At 1.5K the observed decays were all oscillatory whereas at 5K there is a transition from oscillatory to monotonic decay as the energy increases. The sign of the signals inverts near to $E_F$ as discussed in the text.

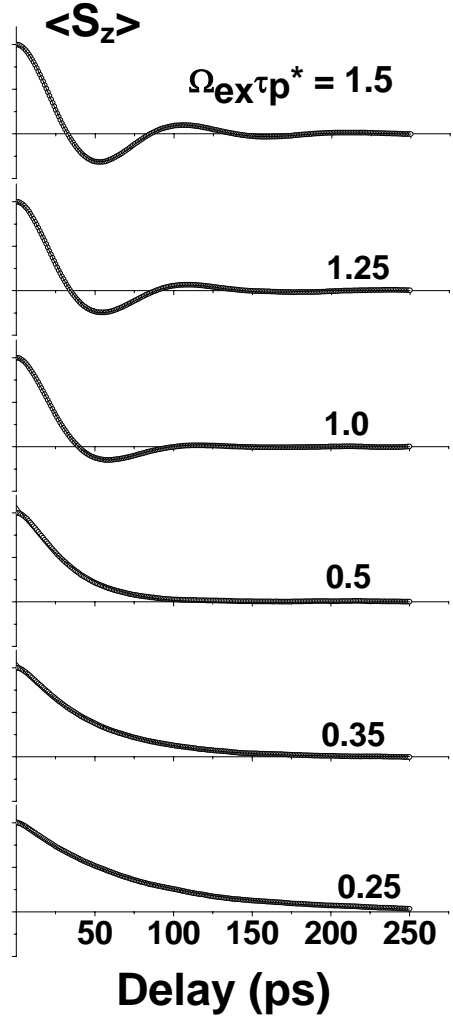

**Figure 2** Results of Monte Carlo simulation of Dyakonov-Perel spin decays as described in the text. The precession frequency, $\Omega_{ex}$, was kept fixed at 0.063 ra/ps and the electron momentum scattering time $\tau_p^*$ was varied to give different values of $\Omega_{ex}\tau_p^*$. The transition from oscillatory to monotonic decay occurs at $\Omega_{ex}\tau_p^* = ½$.

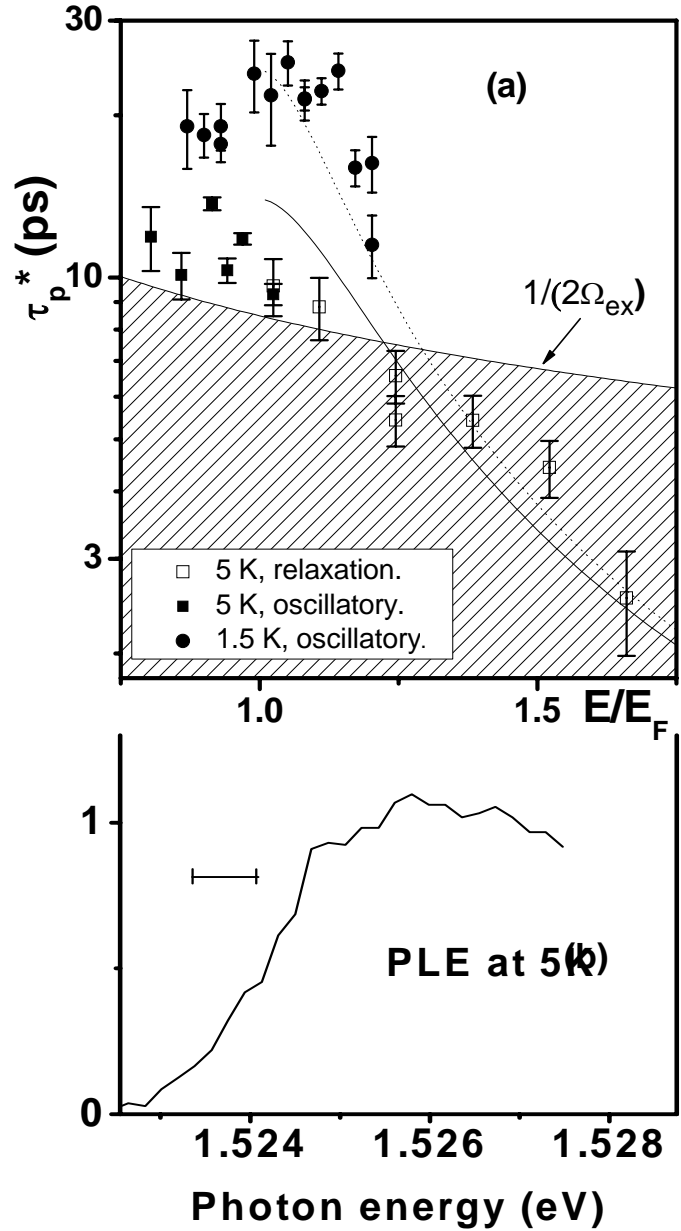

**Figure 3** (a) Experimental values of $\tau_p^*$ extracted from data of figure 1 at 1.5K and 5K using the Monte Carlo simulation method described in the text and in ref 7. Solid symbols correspond to oscillatory spin decays and open symbols monotonic relaxations. Solid and dotted curves are theoretical fits as described in the text. The boundary of the shaded region corresponds to $\Omega_{ex}\tau_p^* = \frac{1}{2}$ in the simulations. (b) Photoluminescence excitation (PLE) spectrum at 5K for detection at 1.516 eV. The horizontal bar indicates the resolution set by the spectral width of the mode-locked Ti-sapphire laser. The Fermi energy $E_F$ was taken to be the half intensity point of the PLE.